\documentclass[elsart1p,letter]{elsart}

\usepackage{epsfig}
\usepackage{rotating}

\def\cf4      {CF${_4}$\ }

\begin{document}

\begin{frontmatter}

\title{
Transport properties of electrons in CF$_4$
}

\author[MIT]{T. Caldwell}
\author[BU,dec]{A.~Roccaro}
\author[MIT]{T.~Sahin}
\author[MIT]{H.~Yegoryan}
\author[LNS]{D.~Dujmic\corauthref{cor}}
\ead{ddujmic@mit.edu}
\author[BU]{S.~Ahlen}
\author[MIT]{J.~Battat}
\corauth[cor]{Corresponding author.}
\author[MIT,LNS,MKI]{P.~Fisher}
\author[MIT]{S.~Henderson}
\author[MIT]{A.~Kaboth}
\author[NLR]{G.~Kohse}
\author[NSE]{R.~Lanza}
\author[MIT]{J.~Lopez}
\author[MIT]{J.~Monroe}
\author[MIT]{G.~Sciolla}
\author[Brand]{N.~Skvorodnev}
\author[BU]{H.~Tomita}
\author[MIT]{R.~Vanderspek}
\author[Brand]{H.~Wellenstein}
\author[MIT]{R.~Yamamoto}

\address[BU]{Physics Department, Boston University, Boston, MA 02215}
\address[Brand]{Physics Department, Brandeis University,  Waltham, MA 02454}
\address[MIT]{Department of Physics, MIT, Cambridge, MA 02139}
\address[LNS]{Laboratory for Nuclear Science, MIT, Cambridge, MA 02139 }
\address[MKI]{MIT Kavli Institute for Astrophysics and Space Research, Cambridge, MA 02139}
\address[NLR]{Nuclear Reactor Laboratory, MIT,   Cambridge, MA 02139 }
\address[NSE]{Department of Nuclear Science and Engineering, MIT,   Cambridge, MA 02139 }
\address[dec]{ Deceased}

\collab{DMTPC Collaboration}

\begin{abstract}
Carbon-tetrafluoride (CF$_4$) is used as a counting gas in particle detectors, but
some of its properties that are of interest for large time-projection chambers
are not well known.
We measure the mean  energy, which is proportional to the diffusion coefficent,  and the attentuation coefficient of electron propagation in CF$_4$ gas
using a  10-liter dark matter detector prototype of the DMTPC project.
\end{abstract}

\begin{keyword}
dark matter\sep directional detector\sep CF4 \sep carbon-tetrafluoride \sep TPC \sep electron diffusion \sep electron attachment \sep DMTPC
\PACS 29.40.Cs \sep 29.40.Gx \sep 95.35.+d
\end{keyword}
\end{frontmatter}

%
%

\section{Introduction}
\label{sec::introduction}

Since carbon tetraflouride (CF$_4$) has a wide spectrum of applications,
most of its properties are well determined (see e.g. ~\cite{Chr04}).
CF$_4$ gas has been suggested as a good target for directional~\cite{spe88},
spin-dependent dark matter searches~\cite{ellis91, Vuilleumier:1993aw, Collar:2000jq, Tanimori:2003xs} 
due to the large spin of the fluorine nucleus and the good ionization-counting properties of the molecule.
However, there is a large discrepancy between existing  measurements of electron transport parameters, 
specifically the diffusion and attachment coefficients, that are of vital importance for the design of 
low-background particle detectors such as dark matter searches.
Detectors that are based on the time-projection-chamber (TPC) concept~\cite{tpc}
 require long drift distances of  ionization electrons through detector gas in a low electric field.
As a swarm of  electrons propagates along an electric  field, it spreads
transversely and longitudinally, which limits spatial precision of the detector.
Current measurements of the transverse spread~\cite{Sch88, Cur88, Lak73}  differ by more than an order of magnitude
 at operating conditions needed for dark matter detection.
Primary electrons can also attach to CF$_4$ molecules (see \cite{Chr04} for details),
which attenuates the electron signal in the detector.
Measurements of the attenuation rate also differ by more than an order of magnitude~\cite{Hun87, Dat92}.
In this work, we present new measurements  of the mean electron energy, that determines the
transverse diffusion, and the attentuation coefficient of electrons in CF$_4$ gas.

%
%

\section{Experimental setup}
\label{sec::setup}

\begin{figure}[hb]
\center
\includegraphics[width=10cm]{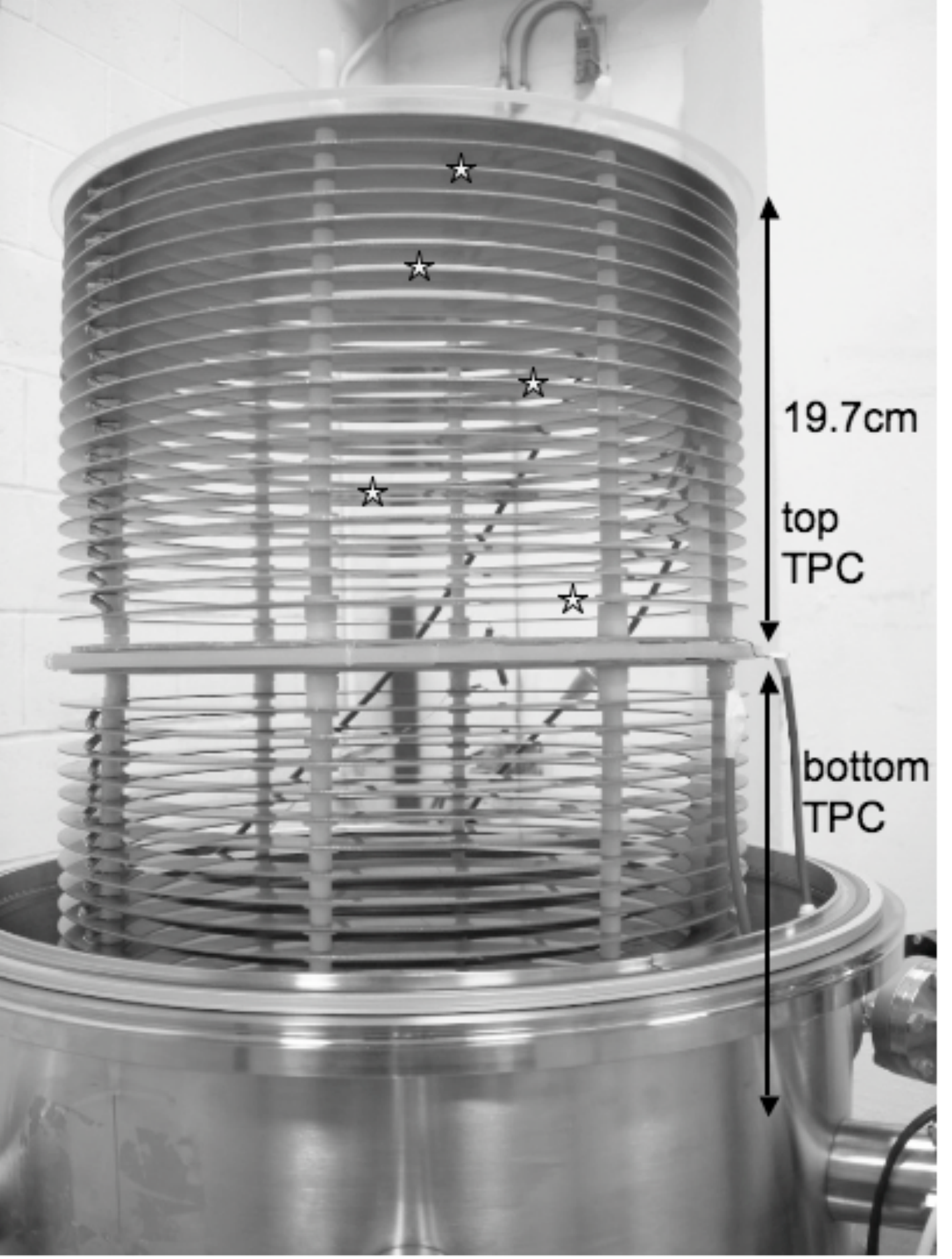} \\
\caption{ Drift cages of the 10 liter DMTPC detector with anode planes in the middle. CCD cameras that image light created
near anodes are placed  above and below the drift cages. The top drift cage is used in measurements.
Positions of alpha sources are marked with stars.
\label{fg::dmtpc10L}}
\end{figure}

The detector (Figure~\ref{fg::dmtpc10L}) is a low-pressure time projection chamber with optical and charge readout.
Particles interacting in the sensitive volume of the detector ionize the gas either directly or
by creating a  charged particle recoil.
The sensitive region of the detector is surrounded by a series of
stainless steel rings with inner diameter of 27~cm, outer diameter of 32~cm, and thickness of 1~mm.
The rings are connected with 1~M$\Omega$ resistors and separated by nylon washers to keep periodicity at 1~cm.
The last ring in the series is connected via a 2.2~M$\Omega$ resistor to a ring holding a grounded
mesh and is separated from the mesh by 1.7~cm. The total height of the drift region is 19.7~cm.
The field uniformity in the drift cage is computed using the finite element method (FEM) with the scaled 
transverse field component, $|E_{\perp}|/|E|$,  less than 1\%.
The cathode is made of a stainless-steel wire mesh with a pitch of 512~$\mu$m and wire diameter of 31~$\mu$m.
The grounded electrode is made of a stainless-steel mesh with a pitch of 256~$\mu$m and wire diameter of 28~$\mu$m.
The detector consists of two  TPCs put back-to-back  such that electrons drift toward
two amplification planes located in the center of the vessel.
Only the top TCP is used in these measurements.

The amplification plane is described in more detail in \cite{Dujmic:2008ut}.
An amplification gap is created by a copper anode (+730~V) 
and a grounded mesh separated with 0.53~mm fishing lines that are placed at 2.5~cm intervals. 
Primary electrons that drift toward the  plane
start avalanches of electrons and scintillation photons~\cite{Pansky:1994zh,Kaboth:2008mi}.  
The detector's primary form of readout is the two-dimensional optical readout of scintillation 
photons created in the avalanche charge multiplication process.  This scintillation light is 
collected with a Nikon photographic lens with f-stop ratio of 1.2 and a focal length of 55 mm.  
The scintillation light is recorded by an Apogee U6 camera with a Kodak 1001E CCD chip.  
In addition to the optical readout, the charge deposited on the anode is also recorded. 
We use a proportional preamplifier (Ortec 109PC) to integrate the
charge signal and a shaping amplifier (Ortec 485) to make unipolar pulses.
A fast digitizer (Alazar ATS860) is used to convert pulses into waveforms,
which are saved with our event record.  The gas pressure is measured by an Inficon PCG400 
pressure gauge which utilizes a combination of a capacitance and a Pirani gauge.

The spatial resolution is estimated by extracting the track width of 5.486 MeV alpha particles from an $^{241}$Am source.
The observed width of the track along the axis in the readout plane is 300-500~$\mu$m.
The resolution is affected by the number of CCD pixels, $n_{bin}$ merged into a readout  bin (98$\mu \rm{m} \cdot n_{bin}$),
the spread of an electron avalanche in the amplification gap ($\sim 80\mu\rm{m}$)~\cite{Biagi},
and the mesh pitch ($256\mu\rm{m}/\sqrt{12} \sim 74\mu\rm{m}$).
The measurement itself is affected by the imperfect collimation of the source (50~$\mu$m)
and the straggling of alpha particles through CF$_4$ gas
(40~$\mu$m)~\cite{srim} in a 1.2~mm-long track segment.
However, the dominant contribution to the spatial resolution uncertainty comes from
the electron diffusion that is described and measured in the next section.

\begin{figure}[hb]
\includegraphics[width=12cm]{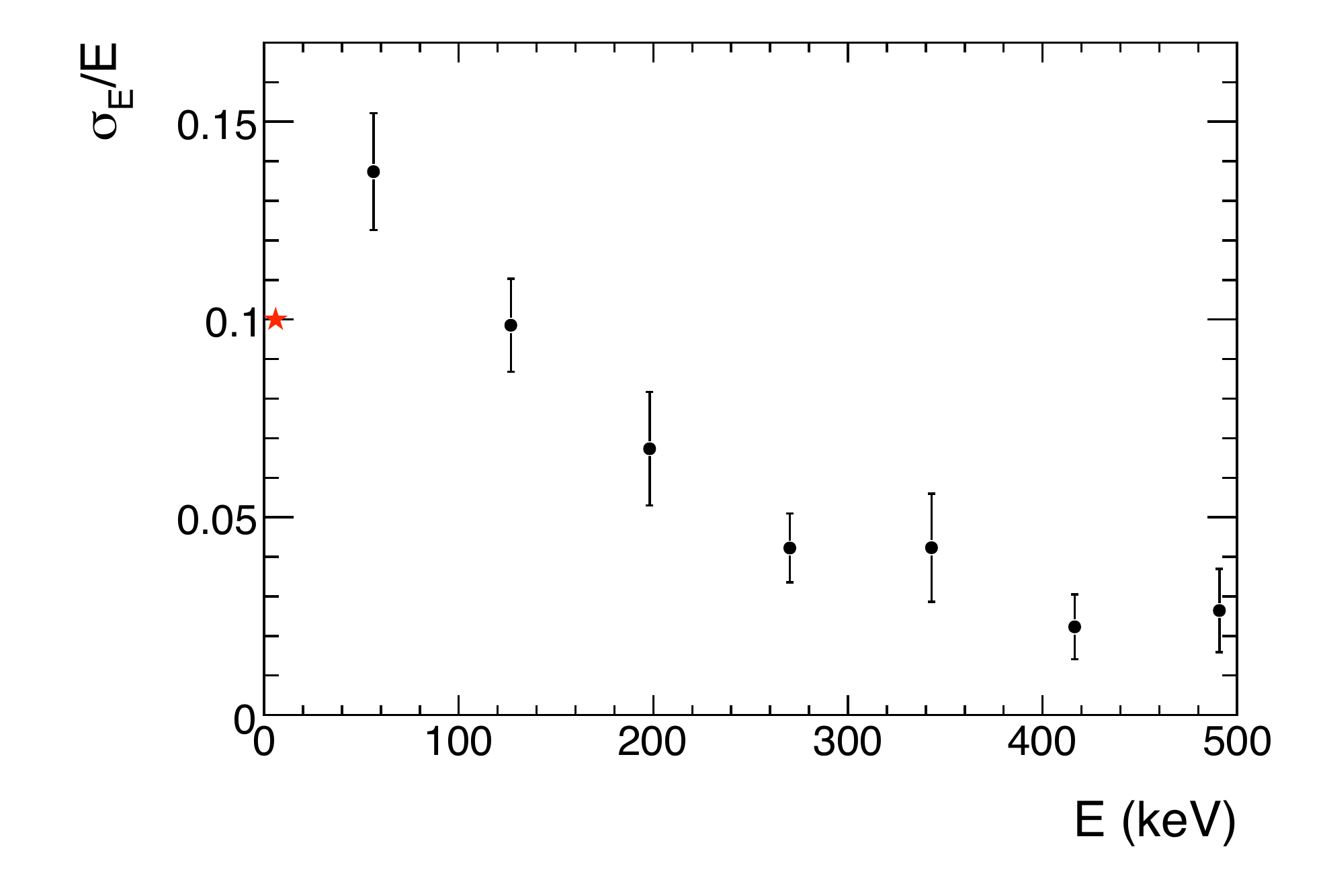}
\caption{Energy resolution measured with the charge and CCD readout.  The star is the $^{55}$Fe peak from charge readout, and the remaining points are from the optical readout of tracks from the $^{241}$Am sources.
\label{fg::e reso}}
\end{figure}

We measure the energy resolution at 5.9~keV for the charge readout using a Fe-55 source. A
background-subtracted spectrum of pulse heights is fitted  with a narrow Gaussian for
the full-absorption peak plus a wide Gaussian for the escape peak.
The resolution is extracted from the narrower Gaussian and is found to be 10\%; a result which is consistent with
Fano factor of one~\cite{Alkhazov:1970fx}.
We infer the energy resolution for the CCD readout from the amount of light that alpha particles
deposit in track segments of
different lengths. The energy deposited into ionization is estimated from the energy loss
and calculated by SRIM~\cite{srim}.
We find that the resolution at 55~keV is approximately 15\%, as shown in Figure~\ref{fg::e reso}.

\section{Measurement of $D/\mu$}
\label{sec::diff}

Electrons propagating through CF$_4$ gas experience a spread in transverse direction
$\sigma^2 \propto D/\mu$ where $D$ is the diffusion constant and $\mu$ is the mobility of
electrons, i.e. a constant of proportionality between the drift velocity and the electric field.
The $D/\mu$ parameter is  a measure of the average electron energy, $\left< \varepsilon \right >= \frac{3}{2}  \left (D/\mu \right )$.

In previous measurements of $D/\mu$, the spread of electron current was determined  with
annular anodes that integrate the swarm current~\cite{Lak73,Hux74}
or with a small-area proportional counter that scans the swarm profile~\cite{Sch88}.
These approaches yield $D/\mu$ values that differ by  more than an
order of magnitude at electric field to number density ratio $E/N\sim 10~\rm{Td}$\footnote{$1~Td=10^{-17}\rm{V~cm}^2$}.
The large data samples of these measurements suggests that the discrepancy is
due to an unknown systematic offset, which motivates a new measurement with a different approach.

\begin{figure}[hb]
\center
\includegraphics[width=10cm]{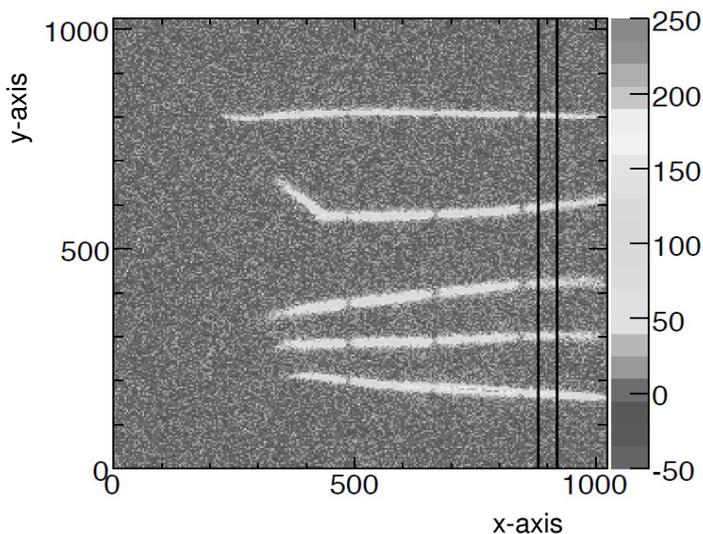}
\caption{A CCD image of 5 active sources of 5.4~MeV alpha particles.
The viewfield is approximately $14.4 \times 14.4$ cm$^2$.
Vertical lines depict a 1.2~mm wide region of interest approximately 3~cm from the sources for measurement of track widths.  
The kink in the second track from the top is likely due to a large-angle scattering on gas nucleus.
\label{fg::imageOfFiveSources}}
\end{figure}

In this work, we extract $D/\mu$ from the width of  alpha tracks ($^{241}$Am) that are
placed at different drift distances inside the drift cage.
A 500~ms CCD exposure with alpha tracks emerging from the five sources is
shown in Figure~\ref{fg::imageOfFiveSources}. The gaps in the tracks are the result of missing light due to the resistive
separators which are used to maintain the amplification gap separation.  Images are taken without
a shutter, so it is possible that some alpha tracks appear during the CCD readout.
Since the CCD is read out in a horizontal direction, these tracks appear shifted
to the left or right. Most of such tracks fall out of the fitting range of interest
or appear at wrong $y$-positions and can be excluded by requiring that the mean
of the light distribution of the track segment agrees with the position of the source.

\begin{figure}[hb]
\center
\includegraphics[width=12cm]{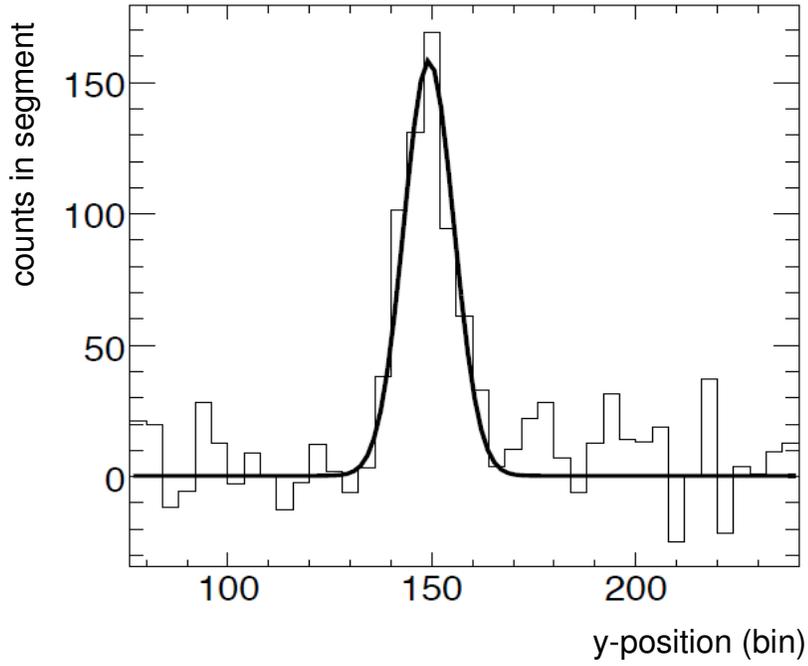}
\caption{A fit to the transverse profile of a track segment for a source at $z=9.1$~cm. 1 bin corresponds to 143 $\mu$m in the image plane.
\label{fg::fitToSource}}
\end{figure}

\begin{figure}[hb]
\includegraphics[width=14cm]{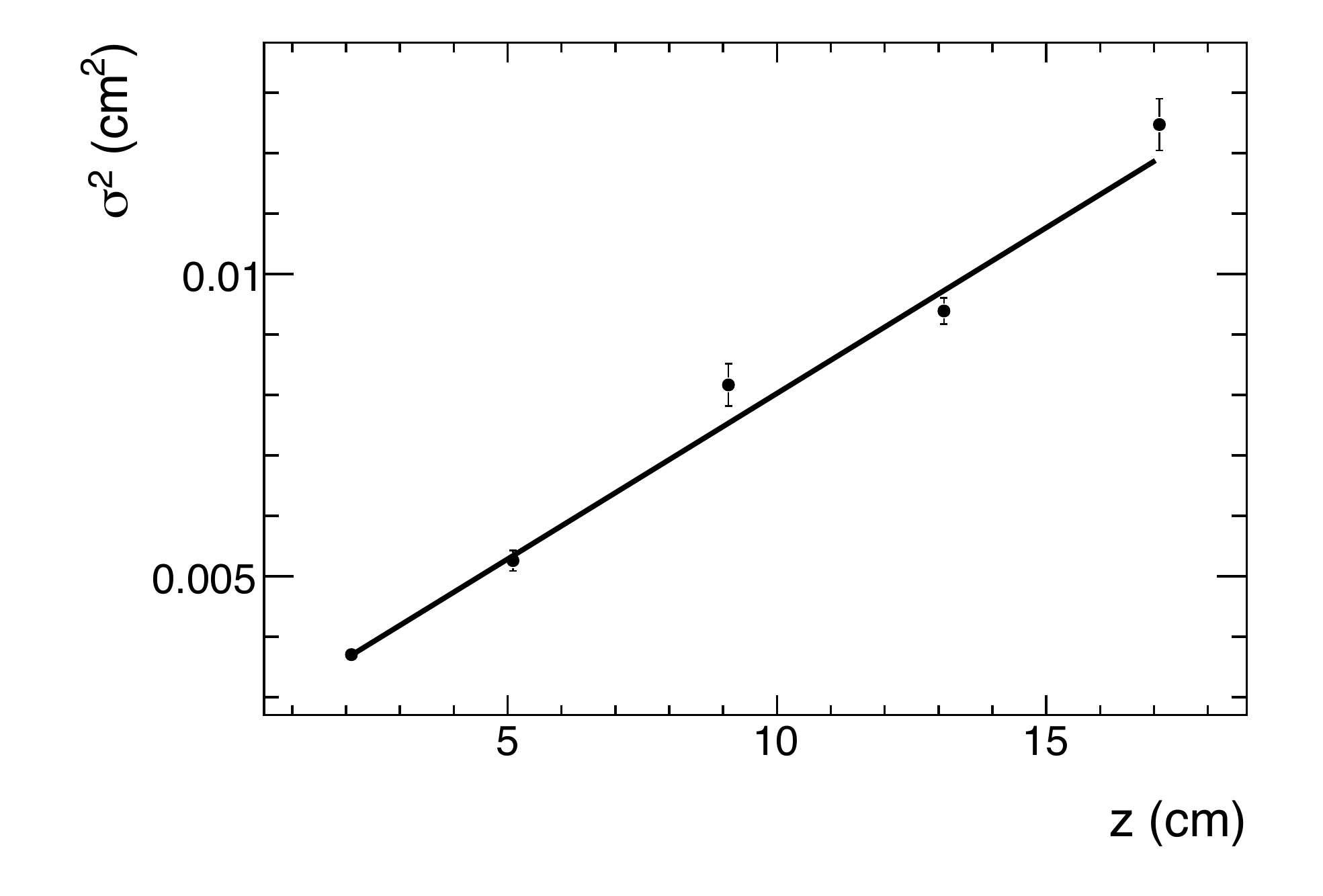}
\caption{Fit of a linear function to $\sigma^2$ versus $z$ for the $E/N=5$~Td data point.
\label{fg::fit}}
\end{figure}

We collect 1000 images with 500 ms exposure for each $E/N$ measurement point. The CF$_4$ pressure is
50-150~Torr, and the temperature is roughly 300~K.
In each track we select a 1.2~mm long segment taken 3~cm from a source and project it onto the $y$-axis.
The projection is then fitted assuming a gaussian signal and a constant term for background, as shown in Figure~\ref{fg::fitToSource}.
We compute the average  width squared, $\sigma^2$, from the collection of all tracks from a single source
that have a yield which is consistent with that of a single alpha track.
The width squared is expected to change linearly with the drift distance, $z$, according to
\begin{equation}
        \sigma^2(z) = \sigma_0^2 + 2 \left ( \frac{D}{\mu} \right ) \left( \frac{z L}{V} \right )
\label{eq::sigma2 experimental}
\end{equation}
where $V$ is the applied voltage in the drift cage of length $L$.
The detector resolution, $\sigma_0$, and $(D/\mu)$ are free parameters that
are determined in a fit to measurement points taken at five different drift distances
and plotted in Figure~\ref{fg::fit}.

\begin{figure}[hb]
\center
\includegraphics[width=14cm]{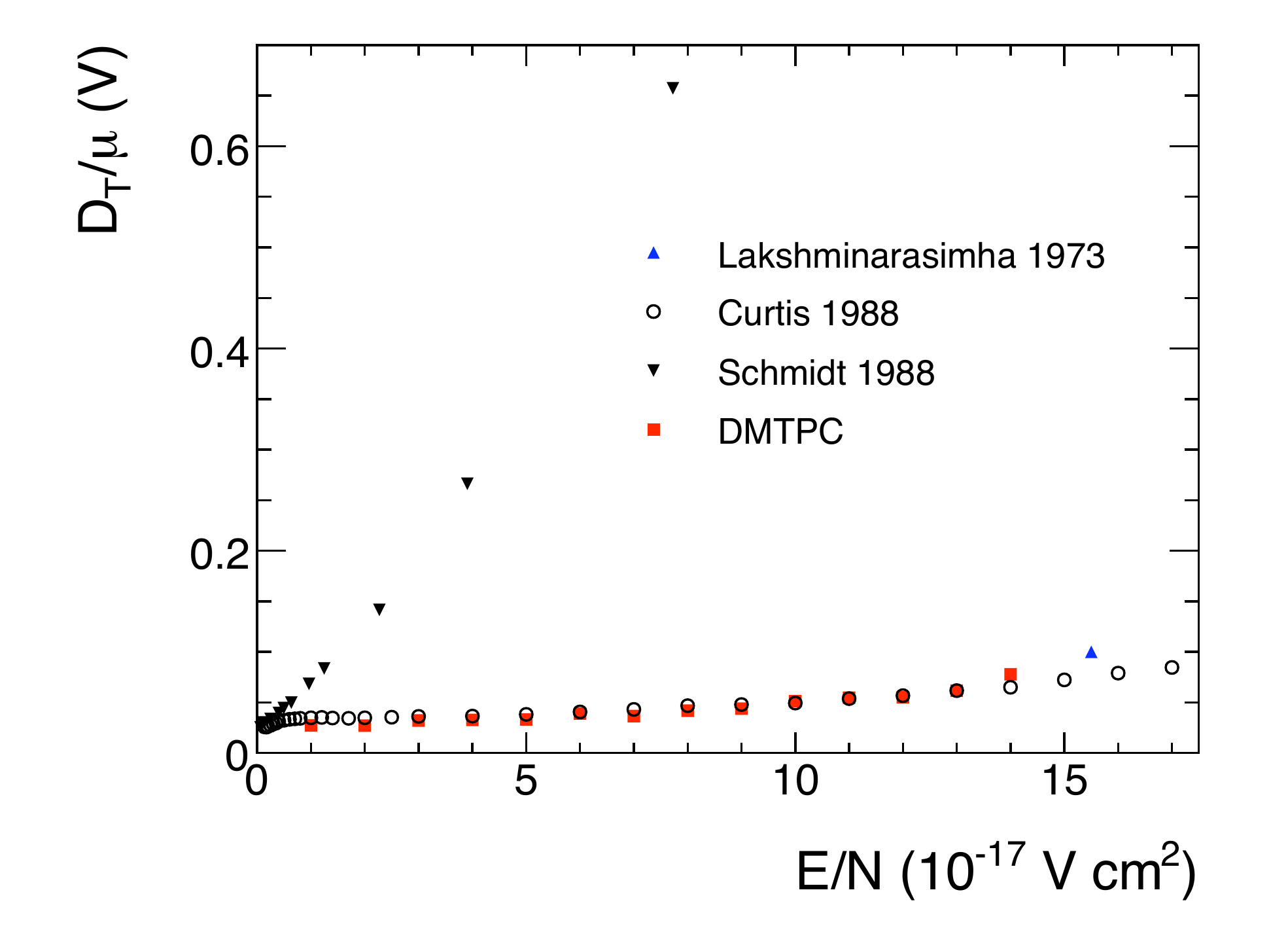} \\
\caption{Measured values of $D/\mu$ as a function of $E/N$ compared with other
measurements.
\label{fg::D/mu result}}
\end{figure}

Data points are taken at three different pressures (50, 75 and 150~Torr) and various drift voltages (1-5~kV).
In regions where  $E/N$ values overlap, measurements are averaged in 1~Td bins of $E/N$.
Results are shown in Figure~\ref{fg::D/mu result} and listed in Table~\ref{tb::D/mu result}.

\begin{table}[t]
\center
\caption{Measured values of $D/\mu$. Errors are dominated by systematic uncertainities that are listed in Table~\ref{tb::syst}. \label{tb::D/mu result}}
\begin{tabular}{r|r}
\hline \hline
$E/N$ (Td)      &       $D/\mu$ (V) \\
\hline
1  &  0.027 \\
2  &  0.027 \\
3  &  0.032 \\
4  &  0.033 \\
5  &  0.033 \\
6  &  0.039 \\
7  &  0.037 \\
8  &  0.042 \\
9  &  0.044 \\
10  &  0.051 \\
11  &  0.055 \\
12  &  0.055 \\
13  &  0.062 \\
14  &  0.078 \\
\hline \hline
\end{tabular}
\end{table}

The dominant source of uncertainty is due to the choice of the position and size
of the  segments used in computation of the track widths.
The dependence on the position of the segment is due to fringe fields in the drift cage that affect
more strongly electrons that travel large drift distances. The percent error due to fringe 
field effects is larger for lower electric field values, as shown in Table~\ref{tb::syst}. 
Also, increasing the length of the segment increases the width of the track due to imperfect collimation
of the alpha sources.
We also assign conservative errors by observing a difference in the $D/\mu$ measurement
after changing the pressure,  the amplification voltage and the quality of focus at the same $E/N$ value.
Gas impurities are taken from the manufacturer's  specification sheet (in ppm):
O$_2$ ($<2$), SF$_6$ ($<1$), N$_2$ ($<4$), CO ($<1$), CO$_2$ ($<1$), H$_2$O ($<1$) and other fluorocarbons ($<2$)~\cite{airgas}.
We confirm that the measurement is immune to outgassing effects by repeating the measurement 
over several days and obtaining consistent results ($\sim1\%$) in both $D/\mu$ and detector gain.
\begin{table}[th]
\center
\caption{Summary of systematic errors for the $D/\mu$ measurement. Contributions are
added in quadrature to give the total. \label{tb::syst} }
\vspace{1cm}
\begin{tabular}{lr}
\hline \hline
   Source                                       &     Error \\
\hline
Collimation, segment size and location          &    11\% ($E/N>5$~Td)   \\
                                                &    20\% ($E/N<5$~Td) \\
Amplification voltage                           &    8.9\% \\
Quality of focusing                             & 4.5\% \\
Gas purity                                      & 5\% \\
CCD readout noise                               & $<1$\% \\
\hline
Total                                           & 16\% ($E/N>5$~Td)   \\
                                                & 23\% ($E/N<5$~Td) \\
\hline \hline
\end{tabular}
\end{table}

\section{Attenuation coefficient}

Attenuation of the signal in a gaseous detector is possible due to attachment of primary ionization electrons 
to CF$_4$ molecules~\cite{Chr04} and  due to pollutants in the detector vessel.
While attachment to impurities like oxygen can be minimized by evacuating the vessel before use and
using a high-purity gas, attachment to molecules of the counting gas itself is unavoidable.
Electron attachment leads to signal attenuation and degraded energy resolution.  
An understanding of the attachment rate is important, particularly for time-projection chambers with large drift distances.  
A signal with initial strength $M_{0}$ will be attenuated to a strength $M$  as a function of drift distance $z$ according to
\begin{equation}
        M=M_0 e^{(\alpha+\eta) z} \approx M_0 e^{\eta z}
\label{eq::eta}
\end{equation}
where $\alpha$ is the electron gain rate ($\alpha \ge 0$) and
$\eta$ is the electron loss rate ($\eta \le 0$).
We assume that the ionization rate $\alpha$ is negligible~\cite{Chr04} so we measure only the attenuation coefficient, $\eta$.
Current measurements of  $\eta$, have large discrepancies: for example, at $E/N=10~\rm{Td}$,   measurements of the loss 
of electrons after drifting 20~cm in 75~Torr of CF$_4$ vary from 0\%~\cite{Hun87} to 70\%~\cite{Dat92}; possibly due to
an unknown systematic offset.

\begin{figure}[hb]
\center
\includegraphics[width=13cm]{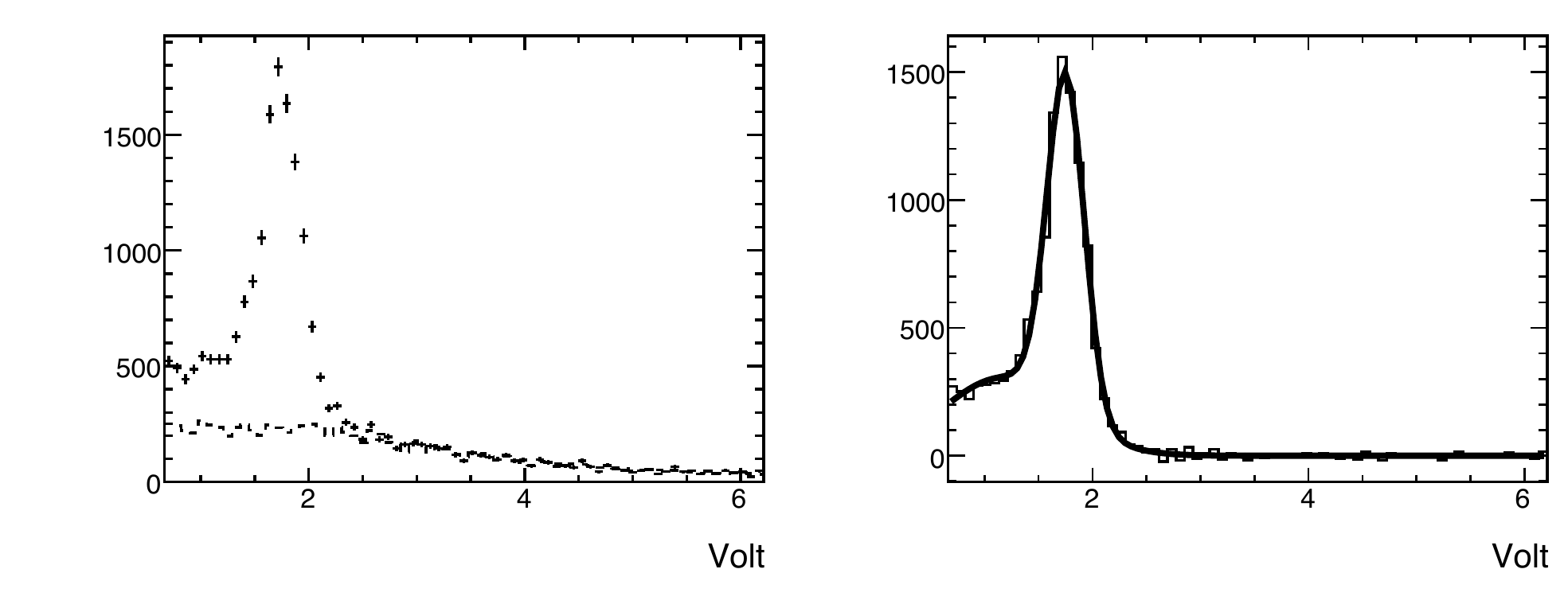}
\caption{ Distribution of pulse heights with an Fe-55  source and
without a source (left). Fit to the background-subtracted distribution (right).
\label{fg::fe55}}
\end{figure}

\begin{figure}[hb]
\center
\includegraphics[width=13cm]{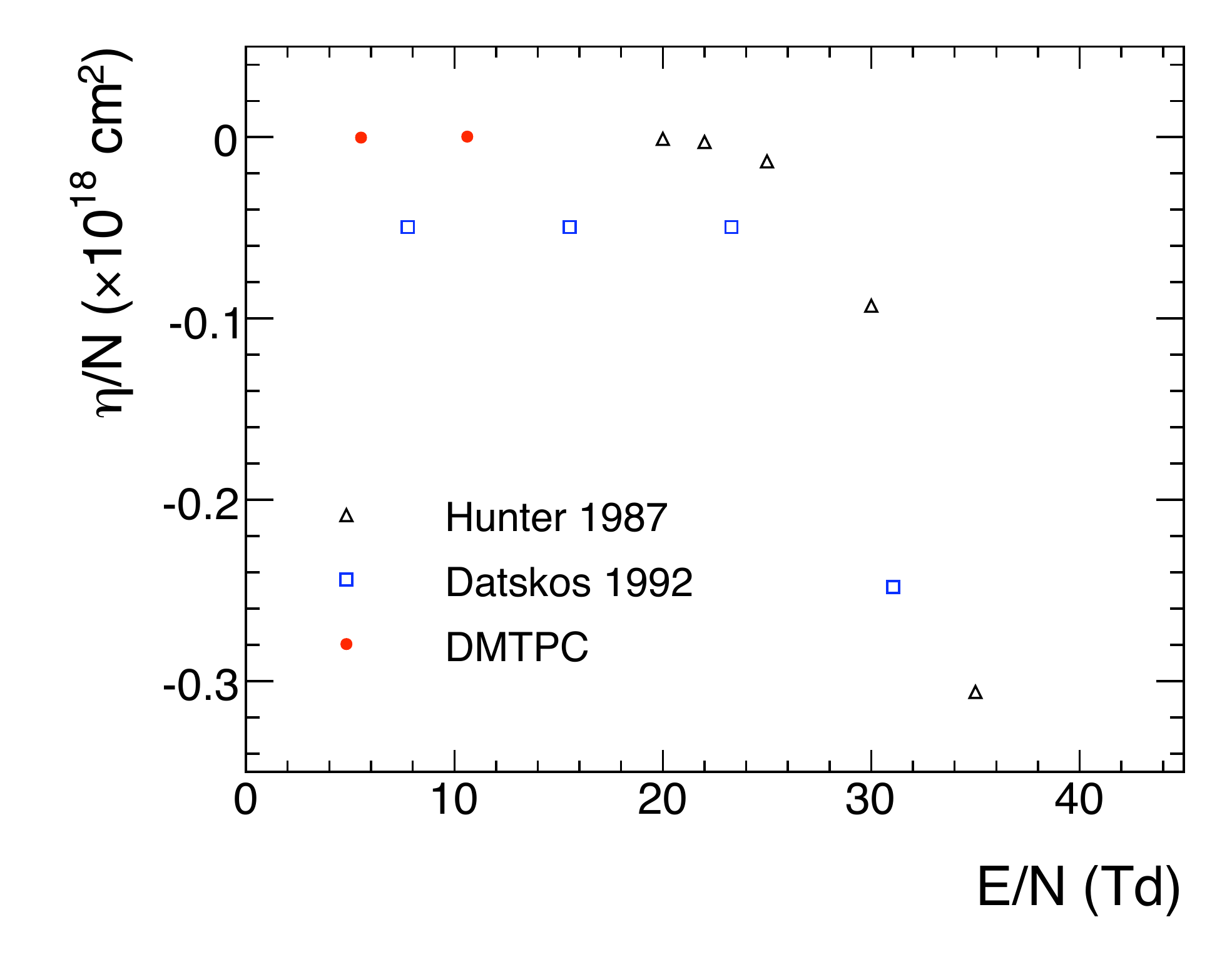}
\caption{Reduced attenuation coefficient ($\eta/N$) as a function of reduced electric field ($E/N$) compared with
previous measurements.
\label{fg::eta}}
\end{figure}

We measure the attachment coefficient by placing a collimated $^{55}$Fe source at different heights and recording
total charge collected at the anode.  The source is located at the edge of the drift volume and placed on a cage 
ring directed radially toward the center of the TPC.  We collect data for 250~sec  with an $^{55}$Fe source  
and 1000~s without any source. Distributions of pulse heights taken with $^{55}$Fe positioned 2.1~cm from the anode
and without any source are shown in the same plot in Figure~\ref{fg::fe55};
both taken at +730~V on the anode and -2600~V on the drift electrode.
In all plots the background distribution is scaled to 250~sec.
In a background-subtracted distribution that is shown in the same figure, we can clearly
identify a 5.9~keV full absorption peak and a smaller, wider escape peak from the same X-ray line,
and we can then describe the distribution with two Gaussians.
The procedure is repeated for different drift heights and voltages.
The attenuation coefficient, $\eta$ in Equation~(\ref{eq::eta}), is extracted by varying the drift height $z$ 
and making measurements of the position of the full absorption peak.
Attenuation parameters at two different values of $E/N$  are listed in Table~\ref{tb::eta}.
\begin{table}[t]
\caption{Measured values of attenuation coefficient in CF$_4$ gas. Systematic errors are estimated to be less than 3\%.\label{tb::eta}}
\begin{center}
\begin{tabular}{r|r}
\hline \hline
$E/N$  (Td)                &      $\eta/N \pm stat$ ($\times 10^{18} ~\rm{cm}^2$)   \\
\hline
5.5               &     $-0.00026 \pm 0.00006$      \\
10.6              &     $0.00027 \pm 0.0001$                 \\
\hline \hline
\end{tabular}
\end{center}
\end{table}
We compare this result with previous measurements in Figure~\ref{fg::eta}, and we find that
and we find that our result is consistent with zero and values from Ref~\cite{Hun87}, but 
inconsistent with the values from~\cite{Dat92}.

The reported value of $\eta/N$ is affected by the uncertainty of the pressure gauge (2\%) and the temperature sensor (2\%).
We make measurements within a few hours of a new fill with CF$_4$ gas,
so the effect of outgasing is expected to be small. This is corraborated with previously mentioned stability in $D/\mu$ and
detector gain over several days.
We confirm that the uncertainty in the spectrum  model has negligible impact on the attenuation measurement
by fitting the spectrum around the full-absorption peak with a single gaussian and observing a change in the peak position of less than 1\%.
We also neglect the error due to imperfect collimation of X-ray source, as this does not change the mean of the drift length.
We add all systematic errors in quadrature.

\section{Conclusion}

We have measured the average  energy ($D/\mu$) and attenuation rate of electrons ($\eta/N$) drifting in CF$_4$ gas
at operating conditions of low-pressure TPC's ($E/N=1-10$~Td).
Discrepancies in previous measurements of these parameters are an obstacle in designing
a large volume TPC that is required for high-sensitivity experiments, such as dark matter searches.
Our measurement of the $D/\mu$ parameter is done by imaging  alpha tracks that are
placed at different heights in the drift cage. The results are consistent with
measurements in Ref.~\cite{Cur88}, but disagree with Ref.~\cite{Sch88}.
The imaging technique can be applied to measurements of $D/\mu$ for other gases and mixtures that allow optical readout.
The systematic errors can be reduced by improving detector uniformity and using smaller CCD pixel bins.

The reduced attenuation coefficient ($\eta/N$) is measured by collecting charge created by  a collimated
X-ray source that is placed at different drift heights. We find that the $\eta/N$ is consistent with zero -- 
a trend predicted by Ref.~\cite{Hun87}, but in disagreement with measurements
in Ref.~\cite{Dat92}.

\end{document}